\documentclass[aps,prl,twocolumn,showpacs,superscriptaddress]{revtex4}

\usepackage{graphicx}
\usepackage{color}
\usepackage{amsfonts}
\begin{document}
\title{Surface states on a topologically non-trivial semimetal: The case of Sb(110)}

\author{Marco~Bianchi}
\affiliation{Department of Physics and Astronomy,
Interdisciplinary Nanoscience Center, Aarhus University, 8000
Aarhus C, Denmark}
\author{Dandan~Guan}
\affiliation{Department of Physics and Astronomy,
Interdisciplinary Nanoscience Center, Aarhus University, 8000
Aarhus C, Denmark} \affiliation{Department of Physics, Zhejiang
University, Hangzhou, 310027 China}
\author{Anna~Str\'{o}\.{z}ecka}
\affiliation{Institut f\"ur Experimentalphysik, Freie Universit\"at
Berlin, 14195 Berlin, Germany}
\author{Celia~H.~Voetmann}
\affiliation{Department of Physics and Astronomy,
Interdisciplinary Nanoscience Center, Aarhus University, 8000
Aarhus C, Denmark}
\author{Shining~Bao}
\affiliation{Department of Physics, Zhejiang University, Hangzhou,
310027 China}
\author{Jose~Ignacio~Pascual}
\affiliation{Fachbereich Physik, Freie Universit\"at Berlin, 14195
Berlin, Germany}
\author{Asier~Eiguren}
\affiliation{Departameto de F\'isica de la Materia Condensada,
EHU/UPV, Barrio Sarriena sn 48940 Leioa, Spain.}
\affiliation{Donostia International Physics Center (DIPC), Paseo
Manuel de Lardizabal, 4. 20018 Donostia-San Sebastian, Spain.}
\author{Philip~Hofmann}
\affiliation{Department of Physics and Astronomy,
Interdisciplinary Nanoscience Center, Aarhus University, 8000
Aarhus C, Denmark} \email[]{philip@phys.au.dk}

\date{\today}

\begin{abstract}
The electronic structure of Sb(110) is studied by angle-resolved
photoemission spectroscopy and first-principle calculations,
revealing several electronic surface states in the projected bulk
band gaps around the Fermi energy. The dispersion of the states
can be interpreted in terms of a strong spin-orbit
splitting. The bulk band structure of Sb has the characteristics
of a strong topological insulator  with a $\mathbb{Z}_2$ invariant
$\nu_0=1$. This puts constraints on the existence of metallic
surface states and the expected topology of the surface Fermi
contour. However, bulk Sb is a semimetal, not an insulator and
these constraints are therefore partly relaxed. This relation of
bulk topology and expected surface state dispersion for semimetals
is discussed.
\end{abstract}

%\pacs{73.25.+i,68.35.Bs}
%\pacs{71.10.-w, 71.70.Ej} % <- the theoretical part

\maketitle
\section{Introduction}
Topological insulators are a recently discovered class of
materials with fascinating properties: while the inside of the
solid is insulating, fundamental topological considerations
require the surfaces to be metallic
\cite{Fu:2007b,Fu:2007c,Moore:2007,Hsieh:2008,Zhang:2008,Moore:2010}.
The metallic surface states show an unconventional spin texture
\cite{Hsieh:2009,Hsieh:2009b} and electron dynamics
\cite{Konig:2007,Roushan:2009,Alpichshev:2010}. They are
furthermore stable in the sense that their existence is a bulk
property and derived from the bulk electronic structure. The
surface state spectrum can be predicted by the single
$\mathbb{Z}_2$ invariant $\nu_0$. For $\nu_0=1$ the bulk
electronic structure is that of a strong topological insulator and
gap-less, stable surface states are expected whereas this is not
the case for $\nu_0=0$ \cite{Fu:2007c,Fu:2007b}.

The topology of the bulk bands does not only permit the prediction
of metallic surface states, it also puts rigorous constraints on
the number of the bands crossing the Fermi energy in certain
high-symmetry directions, and even on the number of closed Fermi
contours encircling high-symmetry points \cite{Fu:2007c,Fu:2007b}.
These topological predictions were found to be obeyed for the
(111) surface of the three dimensional topological insulators
Bi$_{1-x}$Sb$_{x}$ ($0.09 < x < 0.18$)
\cite{Fu:2007b,Hsieh:2008,Hsieh:2009}, as well as Bi$_2$Se$_3$ and
Bi$_2$Te$_3$
\cite{Zhang:2009,Xia:2009,Noh:2008,Chen:2009,Hsieh:2009c}.

The surface electronic structure of the group V semimetals bismuth
and antimony is very similar to that that of the topological
insulator Bi$_{1-x}$Sb$_{x}$. The similarity to the corresponding
Bi surfaces is not surprising since $x$ is quite small. Indeed,
all Bi surfaces studied so far have been found to support metallic
electronic states, in contrast to the semimetallic bulk
\cite{Ast:2001,Agergaard:2001,Hofmann:2005a,Wells:2009}. It has
been suggested \cite{Agergaard:2001} and later shown that these
surface states are split by the spin-orbit interaction
\cite{Koroteev:2004} and the Bi surface states were found to show
some characteristics that were later discussed in connection with
the topological insulators, such as the absence of back-scattering
\cite{Pascual:2004} or the fact that charge density waves cannot
be formed even for nested Fermi contours \cite{Kim:2005b}. The
similarity  Bi$_{1-x}$Sb$_{x}$ to pure Sb is that the gapped alloy
inherits its topological character from Sb ($\nu_0=1$), not from
Bi ($\nu_0=0$). Thus, Sb has the characteristics of a strong
topological insulator while it is a semimetal, not an insulator.
Experimental electronic structure results have so far only been
reported for the  Sb(111) surface
\cite{Sugawara:2006,Hsieh:2009}.

% in this paper.
In this paper we present results from the  electronic structure of
Sb(110). This surface is interesting for two reasons. The first is
the non-trivial bulk topology of Sb.  The second is that, compared
to to the (111) surface, (110) has more distinct so-called
time-reversal invariant momenta (four instead of two) and this
provides the opportunity to study the surface state topology in
more detail.

An interesting question is what the bulk topological
considerations imply for the semimetal surfaces. Strictly spoken,
there is no fundamental reason to expect metallic surface states
on Bi or Sb but such states have so-far always been found and
appear to be quite robust. For Sb(110), we argue that while the
semimetallic character of the substrate inhibits a statement on
the global existence of surface states, the dispersion of states
in certain high-symmetry directions of $k$-space (directions
without bulk projected states) can still be rigorously compared to
topological predictions.

\begin{figure}
\begin{center}
\includegraphics[width=\columnwidth]{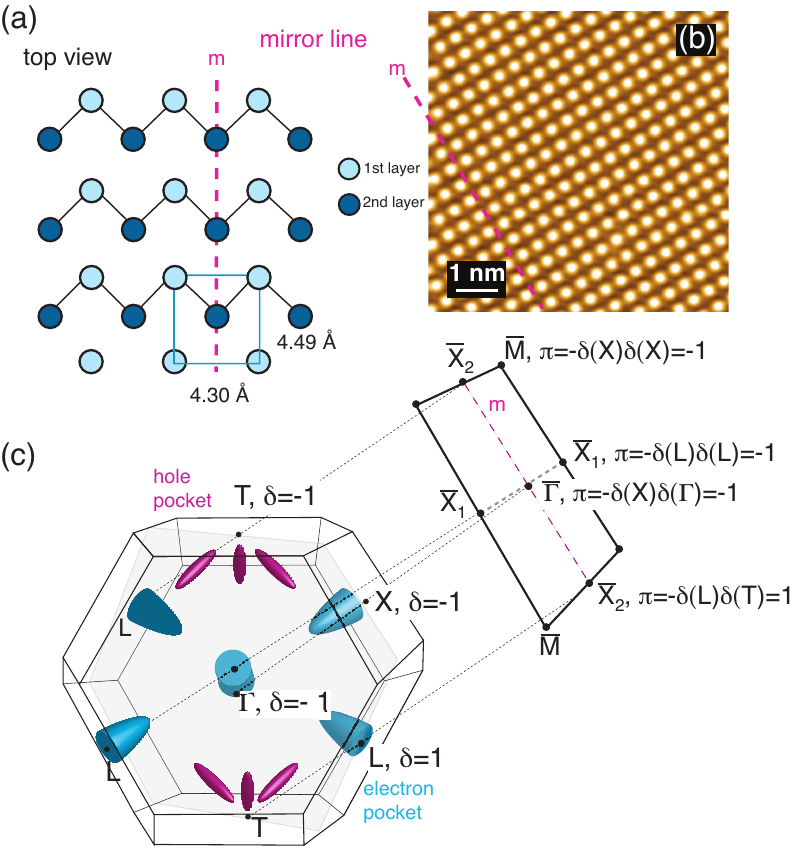}
\caption{(Color online) (a) Truncated-bulk geometric structure of
Sb(110). (b) STM topography of the surface. (c) Bulk Brillouin
zone of Sb with a sketch of the Fermi surface elements
(qualitatively and not to scale), together with a projection onto
the Sb(110) surface Brillouin zone. The grey plane is a bulk
mirror plane which projects onto a surface mirror line. For the  8
bulk time-reversal invariant momenta (TRIMs) ${\Gamma}_i$, the
parity invariants $\delta({\Gamma_i})$ are given. These are
projected onto the 4 surface TRIMs ${\lambda}_a$, resulting in the
surface fermion parity
$\pi(\lambda_a)=-\delta(\Gamma_i)\delta(\Gamma_j)$. The resulting
values of  $\pi(\lambda_a)$ are denoted at the surface TRIMs
\cite{Teo:2008}. \label{fig:1}}
\end{center}
\end{figure}

Figure \ref{fig:1} provides an overview of the Sb(110) surface in
real and reciprocal space.  Figure \ref{fig:1}(a) and (b) show a
model of the truncated bulk surface and the STM topography \cite{Horcas:2007},
respectively. Neither STM nor low energy diffraction give any
indication of a surface reconstruction. Figure \ref{fig:1}(c)
illustrates the bulk Brillouin zone of Sb with the bulk Fermi
surface and the projection of the former onto the (110) surface. A
remarkable feature of the Sb(110) surface is its low symmetry with
a mirror line being the only symmetry element. Nevertheless,
time-reversal symmetry guarantees the electronic structure for
high symmetry points such as $\bar{M}$ to be the same for all four
equivalent points \cite{Agergaard:2001}, effectively giving rise
to a second mirror symmetry in the observed band dispersion.

\section{Experimental and Theoretical Methods}
We have investigated the surface electronic structure of Sb(110)
using angle-resolved photoemission (ARPES). Data were taken at the
SGM-III beamline of the synchrotron radiation facility ASTRID in
Aarhus \cite{Hoffmann:2004}.  The combined energy  and angular
resolution was better than 10 meV and  $0.13^\circ$, respectively.
The data shown here were all taken using a photon energy of 20~eV.
The Sb(110) surface was cleaned \emph{in situ} by cycles of
sputtering and annealing to 520 K. The clean surface showed a
($1\times1$) low energy electron diffraction pattern with the only
symmetry being the mirror line that defines the $\bar{\Gamma}
\bar{X}_2$ direction of surface Brillouin zone \cite{Sun:2006}.
During the photoemission measurements the sample was kept at a
temperature of 60~K. We have also studied the surface topology
using scanning tunneling microscopy (STM) at 5~K.

The Sb(110) surface was modeled considering a repeated slab system
consisting of 54 layers, relaxed up to forces $< 10^{-4}$~Ry/a.u.
We used fully relativistic norm-conserving pseudopotentials as
described in Ref. \cite{Corso:2005}, with the energy cutoff
corresponding to E$_c$=60~Ry. We considered the
Perdew-Burke-Ernzerhof implementation \cite{Perdew:1996} of the
generalized gradient approximation within a non-collinear
implementation of density functional theory (DFT)
\cite{Corso:2005,pwscf}. The self consistent DFT cycle was
completed with a 20$\times$20 Monkhorst-Pack mesh. In order to
calculate the projection of the bulk band structure onto the (110)
surface, we have also used the tight-binding scheme of Liu and
Allen \cite{Liu:1995}. This is expected to give reliable results
very close to the Fermi energy since the tight-binding parameters
have been determined to reproduce the essential features of the
bulk Fermi surface.

\section{Results}

\begin{figure}
\begin{center}
\includegraphics[width=\columnwidth]{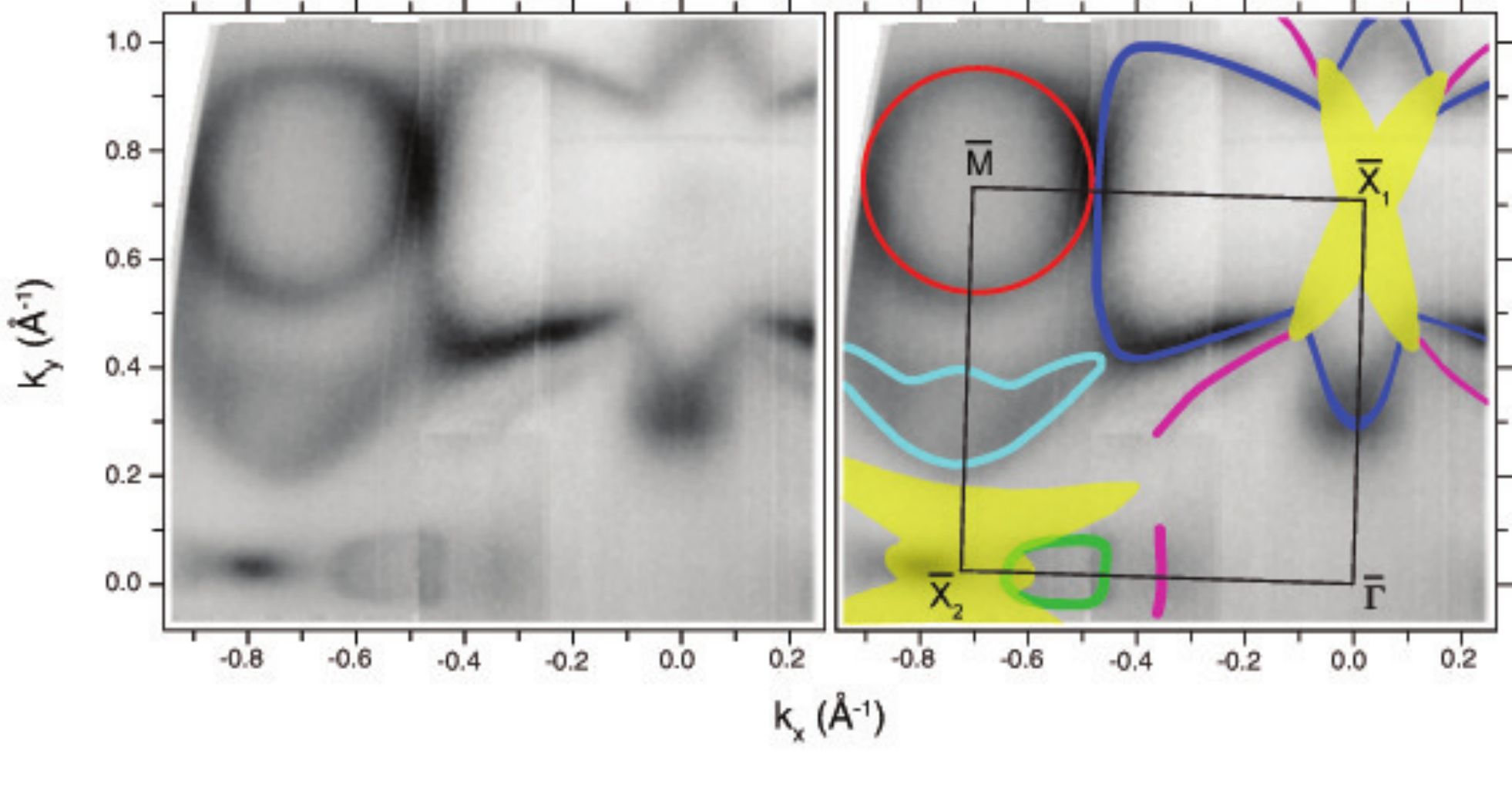}
\caption{(Color online) Photoemission intensity at the Fermi
level. Dark corresponds to high intensity and the dark features
outside the projected bulk band continuum are interpreted as the
surface Fermi contour. The left part of the figure shows the raw
data whereas the different structures are indicated by colored
lines as a guide to the eye on the right part. Different colors
are used for different surface state features. The light yellow
areas correspond to the projected bulk Fermi surface (states
within $\pm5$ meV of the Fermi energy), calculated using the tight-binding
parameters from Liu and Allen \cite{Liu:1995}.
 \label{fig:2}}
\end{center}
\end{figure}

\begin{figure}
\begin{center}
\includegraphics[width=\columnwidth]{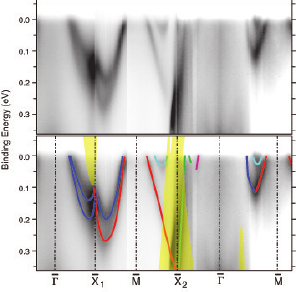}
\caption{(Color online) Photoemission intensity as a function of
binding energy along different directions of the surface Brillouin
zone. The upper part shows the raw data. In the lower part, the
observed bands are emphasized by colored lines. The colors are
corresponding to those used in Figure \ref{fig:2}. The light
yellow areas represent the projected bulk band structure.
 \label{fig:3}}
\end{center}
\end{figure}

The results of the ARPES investigation are shown in Figs.
\ref{fig:2} and \ref{fig:3}. The figures show the photoemission
intensity at the Fermi level and as a function of binding energy
along some high-symmetry directions, respectively. For clarity,
both figures show two versions of the data: only the measured
intensity and this intensity superimposed with colored lines to
guide the eye for identifying the surface states and the projected
bulk band structure. Note that the colored lines are only drawn
where the data shows clearly identifiable structures; they are not
representation of the actual Fermi contour or dispersion that is
expected to continue even if the colored lines end.

The Fermi contour in Figure \ref{fig:2} shows several
surface-related features, identified by their location outside the
projected band continuum and the fact that their position is
insensitive to the photon energy used.  Most pronounced are a
circular contour around  $\bar{M}$ (outlined in red) and a
butterfly-like feature that encloses  the $\bar{X}_1$ point
(blue). Two smaller pockets are seen along the $\bar{M}\bar{X}_2$
line (light blue) and the $\bar{X}_2 \bar{\Gamma}$  line (green).
The latter falls partly into the bulk continuum and thus  has the
character of a surface resonance there. Finally, we find a faint
trace split off the `butterfly wing' and dispersing towards
$\bar{X}_2$ (magenta) and some faint intensity crossing the
$\bar{X}_2 \bar{\Gamma}$ line (also magenta). As we shall see
later from the calculated electronic structures, these features
are probably joint to form a large hole pocket around
$\bar{\Gamma}$, but this is not clearly seen in the data.

The detailed character of the features emerges from the dispersion
shown in Figure  \ref{fig:3}. The circular contour around
$\bar{M}$ (red) is a hole pocket whereas the feature along
$\bar{M}\bar{X}_2$ (light blue) is a shallow electron pocket.The
two features can be interpreted as spin-orbit split partners
stemming from the same state. The state  is unoccupied and assumed
to be spin-degenerate at $\bar{M}$, but split away from this point.
One of the split bands disperses steeply downwards and forms the
hole pocket. The other one forms the electron pocket along
$\bar{M} \bar{X}_2$. At $\bar{X}_1$ the surface state is also
two-fold degenerate but it is occupied and can thus be observed by
ARPES. Away from $\bar{X}_1$ the state clearly splits into two
bands, both along $\bar{X}_1 \bar{M}$ and  $\bar{X}_1
\bar{\Gamma}$, but these bands merge again and are too close to be
distinguished at the Fermi level crossings along these
high-symmetry directions. Consequently, the Fermi level crossings
along  $\bar{X}_1\bar{\Gamma}$ and $\bar{X}_1\bar{M}$ are double
crossings. Close to the  $\bar{X}_1\bar{M}$ direction the two
bands forming the double crossing separate into the circle and the
butterfly.

The weakest features in the data is the  band which  splits off
from the butterfly structure and disperses towards  $\bar{X}_2$
(magenta). Its presence is clearly required by the overall Fermi
contour topology: the two spin-split surface state branches are
occupied at $\bar{X}_1$ and empty at $\bar{M}$ and $\bar{\Gamma}$.
Consequently, two Fermi level crossings have to
be found along the corresponding high-symmetry lines.
Along the $\bar{X}_1 \bar{M}$ direction, the two crossings are formed 
by the circular contour around  $\bar{M}$ and  the `wing' of the butterfly, which is non-degenerate. Along $\bar{X}_1 \bar{\Gamma}$, the blue feature is two-fold degenerate.
The weak magenta feature correspond to the second Fermi level crossing. 
As it disperses away from the butterfly, its intensity diminishes
so much that it cannot be established whether
it continues to the $\bar{X}_2 \bar{\Gamma}$ line and merges with the other
magenta feature observed there, or if it merges into the
projected bulk bands close to $\bar{X}_2$.

The identification of the electronic structure near the $\bar{X}_2
\bar{\Gamma}$ line is more difficult due to the presence of bulk
states. As pointed out above, the feature outlined in green
appears to be a closed pocket around this line but it falls partly
into the bulk continuum. Also, only the band giving rise to the
crossing nearest to $\bar{X}_2$ is clearly identifiable in the
dispersion, the crossing further away from $\bar{X}_2$ is very
weak and the dispersion cannot be followed to higher binding
energies. Nevertheless, the dispersion of the first band suggests that
the pocket is a hole pocket. The magenta feature is well-separated
from the bulk continuum at the Fermi level but it is very broad
and  its dispersion is only clearly observed near $E_F$.  The sign
of its group velocity would be consistent with the feature being
part of a hole pocket around $\bar{\Gamma}$.

\begin{figure}
\begin{center}
\includegraphics[width=0.9\columnwidth]{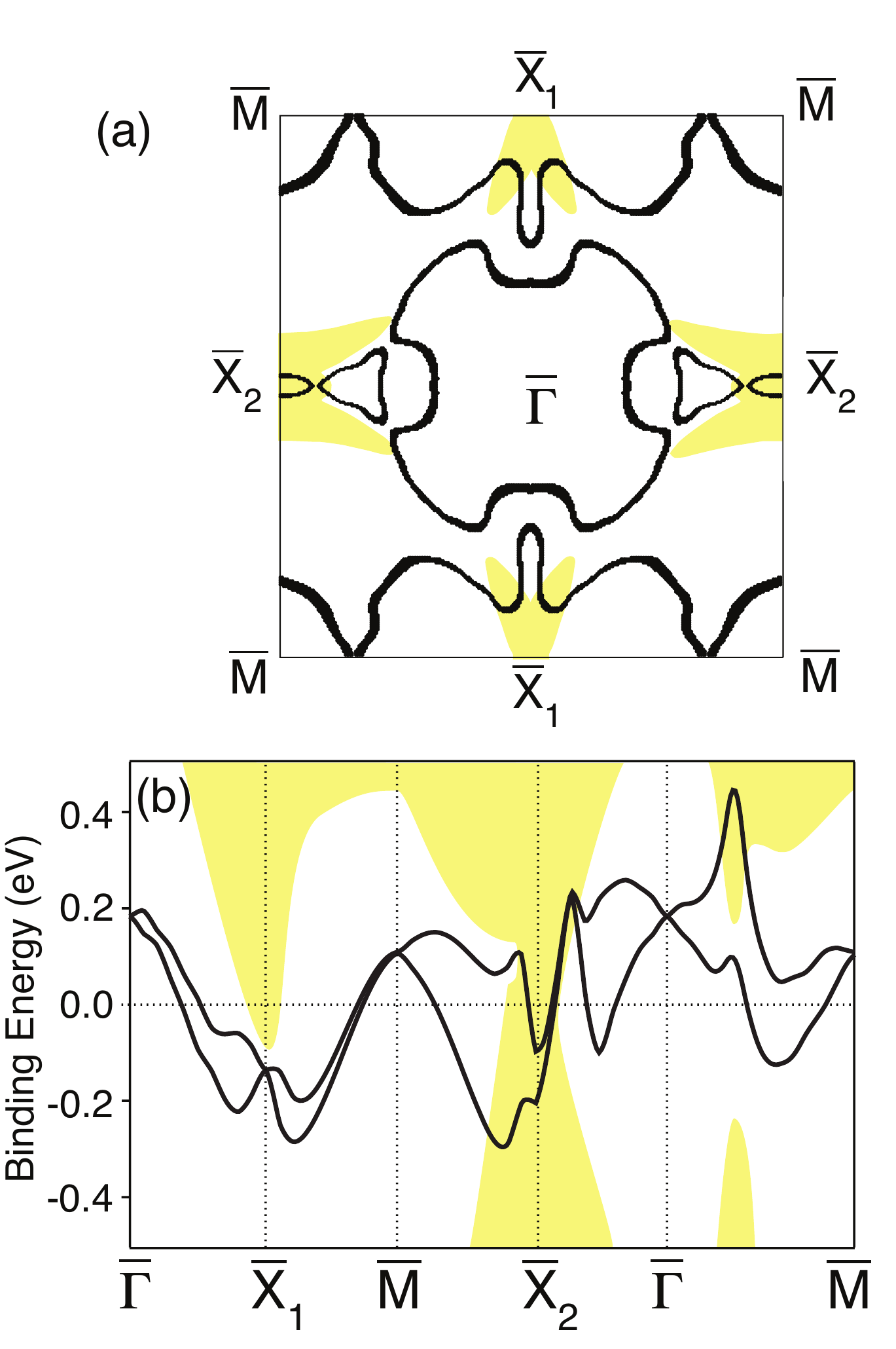}
\caption{(Color online) (a) Calculated Fermi contour and (b)
surface state dispersion. The lines represent the surface states
from the DFT slab calculation. The yellow continuum is the
projected bulk band structure calculated using the tight-binding
parameters from Liu and Allen \cite{Liu:1995}.
 \label{fig:4}}
\end{center}
\end{figure}

The calculated Fermi contour and surface state dispersion is given
in Fig. \ref{fig:4}. Overall, a very good agreement with the
experimental findings is obtained, especially considering the
small energy scale of the bands (a few hundred meV) and the fact that
self-energy effects are not incorporated. In fact, the main
features of the measured Fermi contour are immediately recognized,
especially the butterfly close to $\bar{X}_1$ and the hole pocket
around $\bar{M}$. The calculation also confirms the interpretation
of the surface states as being non-degenerate spin-split bands
that are degenerate only at points where this degeneracy is
enforced by symmetry (so-called time-reversal invariant momenta, see below).

%has been modified
Some smaller details are not entirely captured by the theoretical
results. First, the very shallow electron pocket along $\bar{M}
\bar{X}_2$ appears in the calculations as a dip in the dispersion
of the band, which does not cross the Fermi level. This can easily
be accounted for by small uncertainties in the calculations, like
for example, a slight error in the position of the Fermi level or
in the shape of the band dispersion. The second apparent
difference between experiment and theory is the presence of a
large hole pocket encircling the $\bar{\Gamma}$ point in the
latter. However, a slightly higher Fermi energy in the theory
would cause the hole pocket to merge with the butterfly structure,
giving rise to the experimentally observed double crossing on the
$\bar{\Gamma} \bar{X}_1$ line and to the weak structure split off
the butterfly when going from this line towards $\bar{M}$. In
fact, merging the magenta structures in Fig. \ref{fig:2} would
give rise to a large electron pocket around $\bar{\Gamma}$ and it
seems likely that the lines should be merged since an open Fermi
contour would be unphysical.

As far as the surface state dispersion is concerned, we can thus
draw the following conclusions: the electronic structure of
Sb(110) shows several electronic surface states crossing the Fermi
level. It thus has the character of a good metal in contrast to
the semimetallic bulk. This appears to be a general feature of the
Bi and Sb semimetal surfaces and it has been explained as
resulting from the combination of symmetry breaking and a strong
spin-orbit
interaction\cite{Agergaard:2001,Koroteev:2004,Hofmann:2006,Wells:2009,Sugawara:2006}.
The electronic structure of Sb(110) is similar to that of Bi(110)
in some parts of the surface Brillouin zone. The observed and
calculated dispersion is mostly consistent with the expected
degeneracy at high symmetry points (see below) and a spin-orbit
splitting away from these points. While we do not directly measure
the spin of the bands, this interpretation appears to be based on
solid ground. It is consistent with the findings on all other Bi,
Sb and Bi$_{1-x}$Sb$_{x}$ surfaces and with the calculations
reported here. Moreover, the spin-split nature of the states is
also confirmed by the absence of the characteristic
back-scattering standing waves in our STM investigations
\cite{Strozecka_ToBeSubmitted}. Compared to Bi(110), the
spin-orbit splitting is smaller, as expected. This has the
interesting consequence of concentrating the spin-orbit split
bands in an even narrower window around the Fermi energy,
presumably leading to a higher density of states there, and
possibly opening the option of tuning the electronic structure as
to purposefully move van Hove singularities to the Fermi level
\cite{Ast:2007,Ast:2008}.

\section{Discussion}

% comparison experiment theory and general conclusion of situation with weaker spin-orbit interaction.
We now turn back to the more general question of how much
topological considerations affect the existence of surface states
on semimetal surfaces. We discuss this using the specific example
of Sb(110) but we also address other surface orientations and the
case of Bi.

The topological situation for materials with inversion symmetry in
general and Bi, Sb and  Bi$_{1-x}$Sb$_{x}$ in particular has been
studied in great detail by Teo, Fu and Kane \cite{Teo:2008} and we
summarize some of their results here. The $\mathbb{Z}_2$ invariant
$\nu_0$ which dictates the topological character of the solid is
given by
\begin{equation}\label{nu0}
(-1)^{\nu_0}= \prod_{n=1}^8 \delta_i
\end{equation}
where the  $\delta_i$ are the parity invariants of the eight bulk
time-reversal invariant momenta  (TRIMs) ${\Gamma}_i$, defined by
$-{\Gamma}_i={\Gamma}_i+{G} $ where ${G} $ is a bulk reciprocal
lattice vector. For bulk Sb, the TRIMs are $\Gamma$, T, L and  X.
Their $\delta_i$ values are calculated by
\begin{equation}
    \delta \left ( {\Gamma}_i \right ) = \prod_{n=1} \xi_{2n}\left ({\Gamma}_i\right )
\end{equation}
 where the $\xi_{2n}({\Gamma}_i) = \pm 1$ are the parity eigenvalues of the $2n$th occupied band at ${\Gamma}_i$, obtained from a bulk band structure calculation \cite{Liu:1995,Falicov:1966}. The $\delta_i$ values for the bulk TRIMs are given in Figure \ref{fig:1}(c). The key-difference between Bi on one hand and Bi$_{1-x}$Sb$_{x}$ or Sb on the other hand is  that $\nu_0=0$ for the former and 1 for the latter. This stems from a single change of a parity invariant $\delta(L)$ from $-1$ in Bi to $1$ in the other compounds (see Table I in Ref.  \cite{Teo:2008}).

The bulk parity invariants can now be used to describe fundamental
properties of the surface electronic structure. To do this, the
surface TRIMs ${\Lambda}_a$  are defined such that ${\Lambda}_a=
-{\Lambda}_a+{g}$ where ${g}$ is a surface reciprocal lattice
vector. For these points ($\bar{\Gamma}$, $\bar{M}$, $\bar{X}_1$
and $\bar{X}_2$), the surface state bands must be spin-degenerate,
even in the presence of a strong spin-orbit interaction. For each
surface TRIM,
 the so-called surface fermion parity $\pi_i$ can be determined. Essentially, $\pi_i$ is obtained by projecting out the bulk parity invariants onto the corresponding surface TRIMs (see Figure \ref{fig:1}(c)), using the relation $\pi(\Lambda_a)=-\delta(\Gamma_{i})\delta(\Gamma_{j})$ \cite{Teo:2008}. For instance, $\pi$ for $\bar{\Gamma}$ has to be calculated from the parity invariants of the bulk $\Gamma$ and $X$ points, which are both -1 and hence  $\pi(\bar{\Gamma})=-1$. As mentioned above, the only difference in bulk parity invariants between Bi and Sb is at the $L$ point where $\delta(L)=-1$ and $1$ for Bi and Sb, respectively, and this difference implies change the surface fermion parity at $\bar{X}_2$ from -1 to 1.

The surface fermion parity in Figure \ref{fig:1}(c) can be used to
predict the number of closed Fermi contours around a surface TRIM
or the number of Fermi level crossings between two surface TRIMs.
We start with the latter prediction which is easier to verify
experimentally. The number of crossings has to be zero or even if
the two surface TRIMs have the same parity and odd otherwise.
Based on these rules, Teo, Fu and Kane \cite{Teo:2008} have made
detailed predictions of the qualitative surface electronic
structure of many surfaces. We adopt their result for
Bi$_{1-x}$Sb$_{x}$ to the topologically identical case of Sb(110)
and find that there must be an odd number of Fermi level crossings
between $\bar{X}_2$ and any other surface TRIM and an even number
between two surface TRIMs not involving $\bar{X}_2$.

An inspection of Figs. \ref{fig:2}, \ref{fig:3} shows that this is
the case, despite of the presence of bulk Fermi surface
projections. The situation is clearest for the $\bar{X}_1$ point.
At $\bar{X}_1$ we find the two states to be degenerate, as
predicted, and in both the $\bar{X}_1\bar{\Gamma}$ and
$\bar{X}_1\bar{M}$ direction we find two Fermi level crossings,
also agreeing with the prediction.

%has been modified
For $\bar{X}_2$  the situation is difficult to determine solely on
the basis of our experiments, because the intensity of the surface
bands close to this point is very weak, due to the presence of the
bulk band continuum. At first glance, the number of Fermi level
crossings seems to be as predicted from the topological arguments:
along both directions, $\bar{M}\bar{X}_2$  and 
$\bar{X}_2\bar{\Gamma}$, we can identify one closed contour plus
an extra crossing, and thus an odd number of crossings.

%has been modified
We can also compare the experimentally observed number of closed Fermi contours around
each surface TRIM to the topological predictions based on the
surface fermion parity. A surface TRIM with $\pi(\Lambda_a)=-1$ is
expected to be encircled by an odd number of Fermi contours while
the number is zero or even for $\pi(\Lambda_a)=1$. Thus, we expect
an odd number of contours around $\bar{\Gamma}$, $\bar{M}$ and
$\bar{X}_1$ and an even number around $\bar{X}_2$. For $\bar{M}$
this is fulfilled, as the point is encircled by one hole pocket.
$\bar{X}_1$ is also encircled by only one contour, which is
consistent with $\pi=1$. For $\bar{\Gamma}$ the situation is
unclear, as it depends on the weak feature which is split off the
butterfly. It is likely that this feature connects to the observed
crossing along the $\bar{X}_2\bar{\Gamma}$ line, giving rise to a
circular contour around $\bar{\Gamma}$, as also suggested by our
calculation. For $\bar{X}_2$ the situation is again obscured due
to the projected bands around this point and it is difficult to determine whether this state is encircled by any closed Fermi contour.

%has been modified
The situation becomes clearer when we look at the calculated
electronic structure and test it against the topological
predictions. For $\bar{\Gamma}$, $\bar{M}$ and $\bar{X}_1$ and the
lines between these points, the number of closed contours and
Fermi level crossings are as expected. For $\bar{X}_2$, however, 
the topological predictions appear to be violated: 
the calculations show an additional closed Fermi contour around this point, and
thus the number of Fermi level crossings between $\bar{X}_2$ and
any other TRIM becomes even, not odd as predicted. 
The origin of this discrepancy is that the
states lie within the bulk continuum and are thus not
surface states anymore. In fact, the calculations find that the surface band is mixed with bulk states and penetrates deeply into the bulk of the slab.
This is also responsible for the apparent lifting of the degeneracy at 
$\bar{X}_2$ (thinner slabs give a bigger splitting). 
The deep penetration of surface bands into the bulk breaks down the validity of surface fermion parity rules and, hence, the predictions of  the topological theory, which can only be strictly
applied for insulators.

We address this issue in more detail and ask to what degree
topological arguments can be used to make firm predictions as to
the surface state dispersion on semimetal surfaces. In a simple
picture, the need for an odd number of Fermi level crossings
between two surface TRIMs arises because of their different
surface fermion parity. In an insulator, due to the existence of a
global energy gap around $E_F$, the necessary parity change
between surface TRIMs can only be achieved by surface states. On a
semimetal, on the other hand, a surface-projected bulk state can
be used for this purpose, if a projected bulk Fermi surface can be
found between the two surface TRIMs in question. This argument
implies that the number of $E_F$ crossings between TRIMs without
any projected Fermi surface in between them should be as
topologically predicted. For Sb(110) the only such connection is
between $\bar{\Gamma}$ and $\bar{M}$ and this cut does indeed show
an even number of surface Fermi level crossings, as predicted.
Similar arguments can be made for the $\bar{X}_1\bar{\Gamma}$ and
$\bar{X}_1\bar{M}$ directions. Here there is a bulk Fermi surface
projection (at $\bar{X}_1$) but the surface state dispersion lies
completely outside this projection and consequently it is also in
accordance with the topological predictions. In all these
situations, we only observe an even number of Fermi level
crossings between two TRIMs and thus a topologically trivial
situation. Forcing a semimetal surface to be metallic in a
topological sense would require an odd number of crossings between
two TRIMS without a projected bulk Fermi surface element in
between. There is no obvious candidate for such a situation among
the low-index surface of Bi and Sb.

%has been modified
Concluding, we have presented experimental and theoretical results
on the electronic structure of Sb(110), a non-(111) surface of a
topologically non-trivial material. Along the
directions connecting TRIMs without any bulk Fermi surface
contribution, the observed band dispersion is
in excellent agreement with the predictions of the surface bands topology. The topological arguments become invalid if the projected bulk Fermi surface is present and the surface bands are allowed to mix with the bulk states.

The authors gratefully acknowledge stimulating discussions with
Charles Kane, Hugo Dil and J\"urg Osterwalder, as well as support by the Danish Council for
Independent Research - Natural Sciences, the Deutsche
Forschungsgemeinschaft (STR 1151/1-1) and the Spanish
Ministry of Science and Innovation (FIS2010-19609-C02-00).

%\bibliographystyle{apsrev}
%\bibliography{groupreferences_new,local}

\end{document}